# Regulatory assembly of the vacuolar proton pump $V_OV_1$-ATPase in yeast cells by FLIM-FRET

Stefan Ernst[a], Claire Batisse[b], Nawid Zarrabi[a], Bettina Böttcher[b], Michael Börsch*[a]

[a] 3rd Institute of Physics, University of Stuttgart, Pfaffenwaldring 57, 70550 Stuttgart, Germany
[b] EMBL Heidelberg, Meyerhofstraße 1, 69117 Heidelberg, Germany

**ABSTRACT**

We investigate the reversible disassembly of $V_OV_1$-ATPase in life yeast cells by time resolved confocal FRET imaging. $V_OV_1$-ATPase in the vacuolar membrane pumps protons from the cytosol into the vacuole. $V_OV_1$-ATPase is a rotary biological nanomotor driven by ATP hydrolysis. The emerging proton gradient is used for transport processes as well as for pH and $Ca^{2+}$ homoeostasis in the cell. Activity of the $V_OV_1$-ATPase is regulated through assembly / disassembly processes. During starvation the two parts of $V_OV_1$-ATPase start to disassemble. This process is reversed after addition of glucose. The exact mechanisms are unknown. To follow the disassembly / reassembly *in vivo* we tagged two subunits C and E with different fluorescent proteins. Cellular distributions of C and E were monitored using a duty cycle-optimized alternating laser excitation scheme (DCO-ALEX) for time resolved confocal FRET-FLIM measurements.

**Keywords:** $V_OV_1$-ATPase, assembly, FRET, FLIM, single molecule, DCO- ALEX

## 1. INTRODUCTION

In eukaryotic cells, organelle acidification is achieved by $V_OV_1$-ATPases (or V-ATPases). We study the V-ATPase of the yeast *Saccharomyces cerevisiae*. Its main task is the ATP-powered transport of protons from the cytosol into the vacuole [1]. The V-ATPase is found in the vacuolar membrane of yeast cells and comprises two parts $V_1$ and $V_O$ (see Fig. 1). The membrane-embedded $V_O$ part consists of subunits *a, c, c', c'', d* and *e*. $V_O$ provides the proton channel through the lipid bilayer. Proton pumping is associated with the rotary movements of subunits *c, c', c''* and *d* [2, 3]. The $V_1$ part consists of 8 different subunits named A to H. In the $V_1$ part ATP is hydrolyzed to ADP and $P_i$ sequentially at three catalytic binding sites as in the case of the bacterial $F_oF_1$-ATP synthase [4]. Three subunits A and three subunits B form an alternating hexameric structure in $V_1$. The active ATP binding sites are associated mainly with subunits A [5]. Subunits C, E, G and H form the peripheral stalks and connect the catalytic binding sites in $V_1$ with the membrane-bound $V_O$ part. Subunits D and F are part of the rotor [6] which is mainly driven by the conformational changes in the A subunits of the $V_1$ part during ATP hydrolysis.

The activity of the $V_OV_1$-ATPase is controlled by a reversible disassembly process that is induced by starvation, or by glucose deprivation, respectively [7-9]. In the absence of glucose $V_O$ and $V_1$ disassemble (Fig. 1). The $V_O$ part remains in the membrane, but $V_1$ parts diffuse into the cytosol. After 5 minutes approximately 70 % of all $V_OV_1$-ATPases have dissociated [10]. After adding glucose the $V_OV_1$-ATPases quickly reassemble. The short reassembling time rises the question how this process is organized by the yeast cells.

Here we focus on the distribution of subunits C and E. Subunit C plays a key role in the regulation of $V_OV_1$-ATPase. It has been shown that subunit C can bind ATP at millimolar concentrations and could act as an intracellular ATP sensor [11]. In previous work we observed the dissociation of C from the vacuole and detected a fraction of freely diffusing C subunits in the cytosol of yeast cells upon starvation [12]. However, it remained unclear whether other peripheral subunits of $V_1$ were partly associated with C.

..................................................................................................................................................

* m.boersch@physik.uni-stuttgart.de;  phone (+49) 711 6856 4632; fax (+49) 711 6856 5281; http://www.m-boersch.org/

Important for the glucose-induced re-assembly process is the heterotrimeric RAVE protein complex. RAVE also mediates the biosynthetic assembly of the $V_OV_1$-ATPase in yeast cells. Following glucose deprivation-induced disassembling of $V_O$ and $V_1$, RAVE binds to cytosolic $V_1$ subunits E, G and C [13]. After addition of glucose, the $V_1$ part dissociates from RAVE. E and G likely form three peripheral stalks [14-16] which hold together the two parts of the V-ATPase. Subunit C is a connector between the subunits E and G of the catalytic $V_1$ and subunit $a$ of the membrane-bound $V_O$ (Fig. 1). Binding of subunit C to the RAVE complex is possibly a further regulatory mechanism which controls the activity oft the $V_OV_1$-ATPase in yeast cells. When RAVE binding to subunit C is prevented V-ATPase looses its ability for active proton transport.

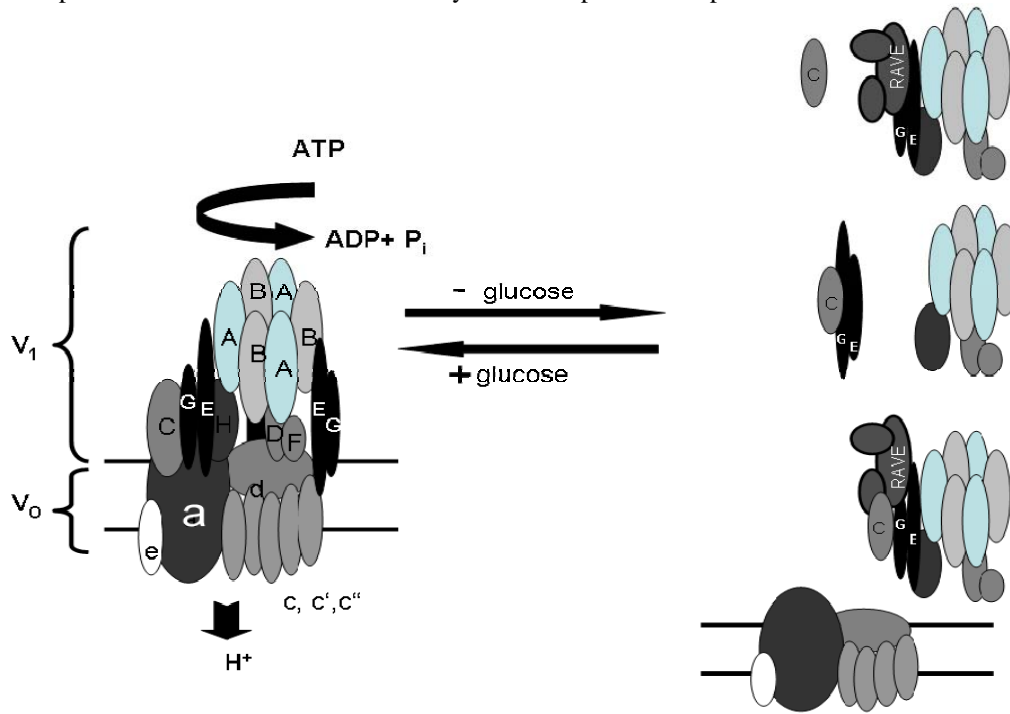

**Fig. 1: Model of glucose-dependent $V_OV_1$-ATPase assembly / disassembly**. The $V_O$ part is embedded in the vacuolar membrane and the $V_1$ part protrudes into the cytosol of the yeast cell. The $V_O$ part consists of the subunits $a$, $c$, $c'$, $c''$, $d$, $e$, and the $V_1$ part of subunits A to H. During glucose deprivation the two parts disassemble. $V_O$ remains in the membrane. The C subunit either diffuses solely into the cytosol or remains associated with parts of $V_1$.

Here we aim at determining whether subunit C forms a complex together with subunits E, G and the RAVE proteins, or whether subunit C diffuses solely in the cytosol with or without the RAVE complex. Therefore we labeled subunits C and E of $V_OV_1$-ATPase with different fluorescent proteins. The yEGFP (yeast enhanced green fluorescent protein) was attached to the C subunit, and mCherry was fused to the E subunits. We apply single-molecule Förster-type fluorescence resonance energy transfer (FRET) to monitor distances and distance changes between subunit C and E after glucose deprivation as well as in the presence of glucose. The distance-dependence of FRET can be used to map the spacing between two intramolecular or intermolecular positions in the range between 2 and 8 nm. For example, the rotary movements of the structurally related $F_OF_1$-ATP synthases have been monitored in real time by single-molecule FRET [17-23]. To image the co-localization of the two subunits C and E we apply a duty cycle-optimized excitation using two pulsed lasers with different wavelengths. Selective photo bleaching of the FRET acceptor is used as a control to prove FRET. In fluorescence lifetime images (FLIM), fluorescence intensities and FRET donor lifetimes indicate the proximity of the FRET-labeled subunits in yeast cells [24-26].

## 2. EXPERIMENTAL PROCEDURES

### 2.1 Preparation of fluorescence labeled yeast cells

For the fluorescence imaging of $V_OV_1$-ATPase in the vacuolar membranes, a number of different mutants of the yeast strain *Saccharomyces cerevisiae* had to be developed. Each mutant contained either one or two fusions of fluorescent proteins two different subunits. Cloning details will be published elsewhere. The mutants were labeled with the yeast-enhanced green fluorescent protein (yEGFP) which acts as the donor fluorophore in FRET imaging. We used mCherry as FRET acceptor fused to the E subunits. To prove FRET in the double mutant we also constructed the FRET donor-only mutant (yEGFP on subunit C).

Yeast cells were grown and stored at 7°C on agar plates. For each imaging experiment yeast cells were freshly grown. Single cell colonies were picked and suspended in 5 ml of YPAD (yeast extract, peptone, adenine and dextrose) medium in a Falcon tube. The solution was softly shaken over night (15-17 h) at 30°C. 100 μl of the cell suspension were mixed with 1 ml of phosphate buffered saline (PBS). The solution was centrifuged for 1 minute at 10000 rpm. The supernatant was discarded and the pellet of yeast cells was resuspended with 1 ml of PBS. The washing procedure was repeated twice to remove the growth medium.

After the last washing step, the pellet was mixed with 5 ml of synthetic complete (SC) medium. The solution was incubated for 4 to 6 h at 30°C. Centrifugation and three washing steps of the pellet with PBS followed. After removal of the SC medium, yeast cells are in a starvation state and $V_OV_1$-ATPase subunits start to disassemble. To reassemble the $V_OV_1$-ATPase on the vacuolar membrane glucose was added (to a final concentration of 3%) and the resuspended pellet was warmed up for 5 Minutes to 30°C. Before imaging, 500 μl of the cell suspension were centrifuged and the pellet was resuspended in 5 μl PBS puffer. Finally 5 μl cells were put on a agarose-coated glass slides to immobilize the yeast cells. Glass slides had been plasma-treated for 5 minutes to smoothen the surface and to remove fluorescent impurities. Subsequently 50 μl of a hot agarose solution in $H_2O$ had been placed on the glass slide. To get a thin film the agarose gel had been deducted with a second slide. After 1 hour drying time, yeast cells could be put on the slide and samples were sealed with a cover slide.

## 2.2 Confocal FLIM-FRET microscope

The custom-built confocal setup is based on an inverted Olympus IX71 microscope (Fig. 2). For the yeast cell imaging, an externally triggered fiber-coupled pulsed laser diode with a wavelength of 488 nm (PicoTa 490, up to 80 MHZ repetition rate, Picoquant, Germany) was used in combination with a continuous wave HeNe laser at 594 nm (Coherent 31-2230-00, 5 mW). The HeNe laser was switched in nanoseconds by an acousto-optical modulator (AOM, model 3350-192, Crystal technologies) for the pulsed duty cycle-optimized laser scheme [27]. Laser beams were expanded with two lenses (f=150 mm and f=50 mm) and overlaid manually using a dichroic beam slitter (DCXR 488, AHF). In epi-fluorescence configuration, laser beams were redirected by a dichroic mirror (either dual band 488/595 or 488/635, AHF, Germany) to a water immersion objective (UPlanSApo 60x W, 1.2 N.A., Olympus) which focused the lasers into the yeast cell layer. Fluorescence passed a 50 μm pinhole and was split into two detection channels by a dichroic beam splitter (DCXR 575, AHF) for FRET imaging. Avalanche photodiodes (AQR14, Perkin Elmer) were used to count photons from yEGFP between 497 nm and 567 nm (interference filter HQ532/70, AHF), and mCherry between 602 nm and 656 nm (629/53, AHF).

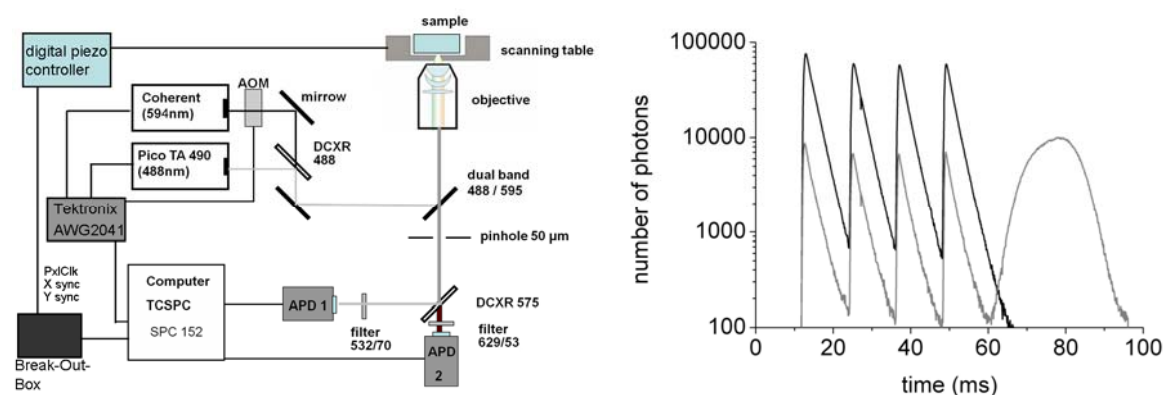

**Fig 2: Confocal microscope for time resolved FRET measurements** of $V_OV_1$-ATPase in yeast cells. Left, living yeast cells were immobilized on an agarose film and scanned with a piezo scanner (range 100 μm x 100 μm). For FLIM-FRET measurements, the TCSPC cards were connected with a BreakOut-Box to receive appropriate trigger pulses. Right, fluorescence decay curves of yEGFP (black trace) and mCherry (grey curve) in the duty cycle-optimized ALEX scheme. Four picosecond pulses at 488 nm in the micro time window between 10 and 60 ns were followed by one AOM-switched pulse at 594 nm.

Single photons were registered by two synchronized TCSPC cards (SPC 152, Becker & Hickl, Germany) and in parallel by a multi channel counter card for online visualization of the intensities and images (NI-PCI 6602, National Instruments). Sample scanning was accomplished by a x-y piezo scanner plus piezo-driven objective positioner (Physik Instrumente). The digital controller was addressed by software written in LabView. The maximal scan range was 100 µm x 100 µm with an average pixel time of 4 ms. Recorded ASCII data were analyzed by a MatLab program "*Yeast-Analyzer*" or by the FLIM software *SPCImage* (Becker & Hickl). FRET analysis was carried out with the software "*Burst-Analyzer*". For fluorescence lifetime imaging the BreakOut-box (SPC-BOB-104, Becker & Hickl) was used. A six-axis digital piezo controller (Physik Instrumente) generated the trigger pulses for an external pixel clock signal and two synchronization signals for the x- and y-scanning directions. Each x-sync pulse started a new line, and each y- sync pulse started a new frame. The BreakOut-Box combines the signal pulses and transfers them to the TCSPC cards operated in the "Scan Sync in" mode. For each pixel of the scanned image, the lifetime was calculated and fitted with a single exponential decay function.

An arbitrary waveform generator (AWG 2041, Tektronix) was used to trigger the pulsed laser, the AOM and the TCSPC cards. The AWG displayed 8 bit digital signals corresponding to an analog waveform with a time resolution of 1 ns. In order to use the AWG as a trigger source for the lasers, ECL output signals had to be transformed into TTL standard pulses. For FRET imaging we used four consecutive laser pulses from the PicoTa 490 laser followed by one pulse of the AOM-switched laser at 594 nm. The four laser pulses were delayed by 10 ns each. The AOM required 30 ns in total for rise and decay. This leads to the smooth shape of the last pulse in the micro time histogram in Fig. 2. With this pulse sequence the maximum excitation rate of the FRET donor was increased compared to a regular alternating laser excitation approach.

## 3. RESULTS

Yeast cells with one or two different fluorescent proteins fused to subunits C or E of $V_OV_1$-ATPase were grown under regular conditions indicating no deleterious effects of the added fluorescent proteins. Images of living yeast cell monolayers during starvation are shown in Fig. 3. Therefore cells were immobilized in thin agarose films and images were obtained in sample-scanning mode by moving with a two-dimensional piezo scanner through the fixed laser foci. Stacks of images at different heights could be recorded using a piezo-driven objective positioner. We applied a duty cycle-optimized alternating laser excitation scheme for FRET imaging and an additional FRET acceptor check or acceptor photo bleaching, respectively (488 nm and 594 nm). Picosecond pulses at 488 nm allowed for fluorescence lifetime measurements, i.e. FLIM. In this microscope configuration, multi channel time trajectories of intensities and lifetimes can be recorded during photo bleaching at selectable positions within a cell. In addition, fluorescence correlation functions (FCS) and cross-correlation functions can be calculated from the same data set of single photons. We focused either into the cytosol or on the vacuolar membranes of the cell to identify the effective hydrodynamic radius of subunit C, i.e. diffusing solely or bound to other subunits of $V_OV_1$-ATPase (data not shown).

$V_OV_1$-ATPase disassembled during starvation of the living yeast cells. Subunit C was mainly distributed in the cytosol (Fig. 3A, 3E) indicated by medium fluorescence intensities. However, in some cells subunit C was still found on the vacuolar membrane resulting in dark rings due to high fluorescence intensities (Fig. 3E in the absence of a mCherry-labeled subunit). FRET efficiency $E_{FRET}$ was calculated for each pixel after correction for spectral bleed through, direct excitation of the FRET accepter (if present), and detection efficiencies according to filter spectra and spectral detector sensitivity of the photon counting APDs. In the presence of mCherry-labeled subunits E, fluorescence intensities in the FRET acceptor image (Fig. 3B) were lower than in the FRET donor image yielding $E_{FRET}$ values between 0.2 and 0.5 (Fig. 3C). Only one cell exhibited higher FRET efficiencies which were located on the vacuolar membrane. Most of the corresponding FLIM lifetimes of the FRET-labeled $V_OV_1$-ATPase in the cytosol were shorter than 2 ns (Fig. 3D).

The control yeast strain contained only the fusion of yEGFP to the C subunit. Fluorescence intensities in the FRET donor image (Fig. 3E) were partly higher compared to the FRET donor image in the presence of an acceptor. However, these high intensities were seen only in cells with subunit C remaining on the vacuolar membrane. The fluorescence intensities detected on the "red" channel in these images (Fig. 3F) were significantly lower than in the respective image shown in Fig 3B. Therefore the calculated apparent FRET efficiencies were low (Fig. 3G). Correspondingly, the FLIM data of subunit C in the cytosol with mean lifetimes longer than 2 ns indicated the absence of resonance energy transfer to other fluorophores on associated subunits.

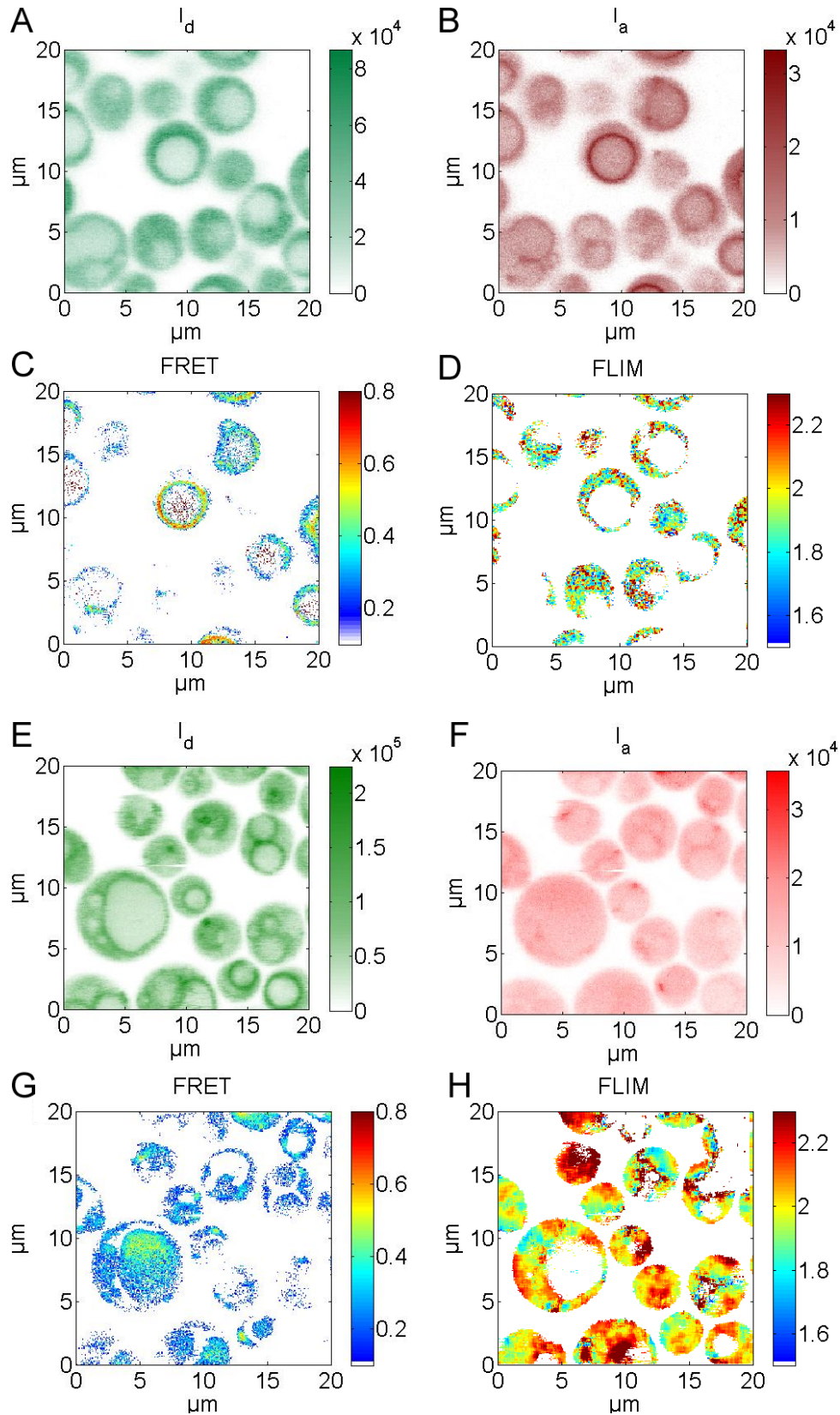

**Fig. 3: FLIM-FRET imaging of living yeast cells during starvation.** A to D, FRET-labeled $V_oV_1$-ATPase. E to H, C-only labeled $V_oV_1$-ATPase. **A**, intensities of FRET donor yEGFP on subunit C. **B**, intensities of FRET acceptor mCherry on E. **C**, calculated FRET image. **D**, fluorescence lifetime of FRET donor yEGFP on subunit C (time in ns). **E,** intensities of yEGFP on subunit C in the absence of mCherry ("donor-only"). **F**, intensities of yEGFP on subunit C detected in the "red" channel due to spectral bleed through. **G**, apparent FRET efficiency in the absence of mCherry due to spectral bleed through. **H**, fluorescence lifetime image of yEGFP on subunit C (time in ns).

## 4. DISCUSSION

The FLIM-FRET data presented here demonstrate the applicability of time resolved imaging of living yeast cells to monitor the variable subunit composition of $V_OV_1$-ATPase parts. $V_OV_1$-ATPase shows a remarkable dis- and re-assembling process in response to nutrition conditions. In order to gain detailed information about this process we focused on subunits C and E of the $V_1$ part. Different fluorescent proteins were genetically fused to these subunits serving as FRET donors or acceptors, respectively. These fluorophores can either be used to probe subunit associations directly by FRET, by diffusion measurements using FCS, or by co-localization using dual-color laser excitation and FCCS.

Previously we showed that the diffusion times of subunit C in the cytosol during starvation indicate a small hydrodynamic radius [12] which corresponds to a freely diffusing subunit without association to other subunit of $V_1$. In contrast, the FLIM-FRET images of double-labeled $V_1$-ATPase provide evidence for Förster-type resonance energy transfer from the fluorophore on C to an attached subunit with FRET acceptor, i.e. binding to subunit E in the cytosol. However, low FRET efficiencies and only a small effect on the fluorescence lifetimes of the yEGFP fluorophore can be interpreted as a partial or a transient binding of C and E in the cytosol. Both types of interaction measurements suffer from the fast photo bleaching of the fluorophores. Diffusion times will be shortened by photo bleaching. Thus the FCS-based calculations of the apparent mean hydrodynamic radius will yield too small values, and an interaction with other subunits is not recognizable. In addition, FCS measurements do not provide information about immobilized fluorophores, for example, if subunit C is bound to actin filaments within yeast cells. *In vitro* binding of subunit C to E and G in a well-defined 1:1:1 stoichiometry strongly support the possibility of subunit interaction [15].

Accordingly, FLIM-FRET measurements might be more reliable. Intensity-based FRET calculations as well as FLIM indicate that only a fraction of subunit C is still associated with subunit E in the cytosol during starvation. However, for the FRET-labeled $V_OV_1$-ATPase we noticed that the FRET acceptor mCherry is bleaching much faster than the donor yEGFP. Measuring time trajectories of fluorescence intensities at positions on the vacuolar membrane in the presence of glucose support that fast photo bleaching of the FRET acceptor can be used to identify FRET at the beginning of the time trajectory (data will be published elsewhere). FRET acceptor mCherry can be excited directly with 488 nm causing photo bleaching. This might explain the lack on any detectable cross-correlation amplitude using alternating lasers for the co-localization imaging and FCCS of subunit C and E in the cytosol.

We also faced the problem of an unstable re-assembly state of $V_OV_1$-ATPase. During the imaging process by slowly scanning the yeast cells through the laser foci, the optimal living conditions after addition of glucose have to be preserved on the microscope. However, we noticed that only one image scan was possible corresponding to an imaging time of about 5 minutes (256 x 265 pixel with 4 ms integration time per pixel). In the subsequent scan, subunit C already dissociated in part from the vacuolar membrane. In order to prolong the measurement time a new heating chamber has been constructed which will maintain the optimal temperature for the yeast cells during the scanning process (30°C).

Finally the spatial resolution of the confocal excitation is limiting. During starvation, the vacuole of the yeast cell expands to a maximum size and the remaining cytosolic volume is minimized. Other organelles like mitochondria or the endoplasmic reticulum are concentrated, and the probability of non-fluorescent diffusion barriers or obstacles within the laser focus for FRET or FCS measurements significantly increases. Optical superresolution techniques like *stimulated emission depletion* (*STED*) to reduce the spot size of the excitation/detection volume beyond the diffraction limit [28], or super localization microscopies like *PALM* based on single fluorophore on-off switching [29] might be required for future experiments addressing the question of the regulatory assembly / disassembly of the vacuolar proton pump $V_OV_1$-ATPase.